# Retrodictive quantum optical state engineering

K. L. PREGNELL and D. T. PEGG*

Centre for Quantum Computer Technology, School of Science,
Griffith University, Nathan, Brisbane, 4111 Australia

**Abstract.** Although it has been known for some time that quantum mechanics can be formulated in a way that treats prediction and retrodiction on an equal footing, most attention in engineering quantum states has been devoted to predictive states, that is, states associated with a preparation event. Retrodictive states, which are associated with a measurement event and propagate backwards in time, are also useful, however. In this paper it is shown how any retrodictive state of light that can be written to a good approximation as a finite superposition of photon number states can be generated by an optical multiport device. The composition of the state is adjusted by controlling predictive coherent input states. It is shown how the probability of successful state generation can be optimized by adjusting the multiport device and also a versatile configuration that is useful for generating a range of states is examined.

## 1. Introduction

The usual formulation of quantum mechanics is predictive. States are assigned to a system on the basis of knowledge of preparation events. These states can then be used for predicting the outcomes of a future measurement. Quantum mechanics, however, also allows retrodiction, that is, the calculation of probabilities of past preparation events on the basis of the known outcome of a measurement [1]. Formulations of quantum mechanics that are symmetric in preparation and measurement have been given in terms of pure states and von Neumann measurements [2] and recently extended to more general state preparation and measurement [3]. Various applications have also been studied [4]. In these formulations a retrodictive state, which propagates backwards in time, can be assigned on the basis of a measurement event. Projection of this retrodictive state onto a preparation device operator at an earlier time yields, with suitable normalization, the probability that the system was prepared in the corresponding state. An immediate application of this is in quantum communication where the recipient of the outcome of a measurement of a quantum system must try to determine the state in which the system was prepared [5].

Although retrodictive states travel backwards in time, they cannot be used to send controllable messages into the past. Essentially this is due to our lack of control over the particular measurement outcome that determines the retrodictive state. Such lack of control also occurs in preparing predictive states by means of a von Neumann measurement, which leaves the system in the measured state after the measurement. The outcome of the measurement is uncontrollable but in the predictive case, because the state generated propagates forwards in time, the experiment can be aborted if the measurement outcome is not the one required. In the retrodictive case it is always too late to abort; one can only exclude a particular case from the final statistics. Despite this, retrodictive states do have other uses in addition to quantum communication. One application is the measurement of quantum optical observables, for example, by projection synthesis [6, 7]. This procedure allows the probability distribution of any observable for any state of light to be measured by means of the projection of this state onto a specially tailored retrodictive state. In the projection synthesis method of [6] and [7] the retrodictive state is prepared as follows. When photodetectors in the two output ports of a beamsplitter register a specified pair $(n_1, n_2)$ of numbers of photocounts a retrodictive state, which is the product of corresponding photon number states $|n_1\rangle_1|n_2\rangle_2$, is produced. This state travels backwards in time through the beamsplitter and becomes entangled. This entangled state is projected onto a specially prepared predictive state in the first input port, resulting in an appropriately engineered retrodictive state in the second input port. If this retrodictive state is, for example, a phase state, then its projection onto an unknown predictive state in the second input port provides a measure of the probability that the unknown state has that particular phase. This particular procedure for engineering a retrodictive state clearly relies on the ability to engineer the appropriate predictive state in the first input port. Retrodictive states can also be used as probes to measure directly the mean, variance and higher moments of quantum optical phase and also for the direct measurement of individual elements of the optical density matrix [8]. These retrodictive probe states are simple number state superpositions and can be generated from coherent states inputs.

Another application of retrodictive states is for predictive state engineering: a one-mode retrodictive state projected onto a two-mode entangled predictive state can result in the generation of an unentangled predictive state in the other mode. A simple example of this is the quantum scissors device used for truncating predictive coherent states to form optical qubit states [9]. The procedure is similar to the beamsplitter method described above with $n_1 = 0$ and $n_2 = 1$ and with a coherent predictive state in the first input port. This time, however, the resulting retrodictive state is allowed to travel backwards in time to one output of a second beamsplitter which has been used to entangle a zero-photon and a one-photon predictive state. The projection onto this entangled predictive state results in the generation of the desired predictive qubit state.

Although it is worth exploring more potential uses for retrodictive states, in this paper we shall concentrate on the general problem of generating an arbitrary retrodictive state with a lossless multiport device comprising beamsplitters, mirrors and phase shifters. Such a device will have the same number, $N+1$ say, of input ports as output ports. The scheme we shall adopt is illustrated in figure 1. There is a photodetector in each output port and in each of $N$ input ports there is a predictive state. The retrodictive state that is the product of the $N+1$ number states corresponding to the photocount numbers from the $N+1$ detectors propagates backwards through the device and becomes entangled. Projection of the $N$ predictive states onto this entangled state results in the required retrodictive state propagating backwards in time from the free input port. To make the scheme as practical as possible, in this paper we shall restrict the predictive input fields to be in easily prepared and controllable coherent states. Photodetector inefficiency will cause the retrodictive field to be in a mixed, rather than a pure, state. To reduce such inefficiency as much as possible, we shall assume that the detectors need only distinguish between one and zero photocounts in the detection period.

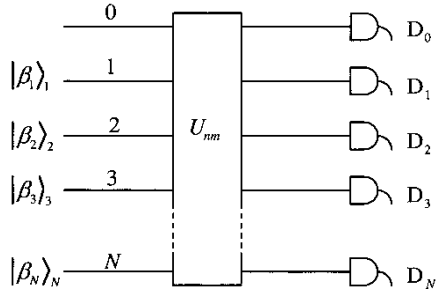

Figure 1. Schematic diagram of multiport device for generating retrodictive states of light. There are $N+1$ input modes and $N+1$ output modes. In input modes $1, 2, \ldots, N$ are predictive coherent states with adjustable intensities and phases. In output modes $0, 1, \ldots, N$ are photodetectors $D_0, D_1, \ldots, D_N$. If the coherent states are appropriately adjusted the retrodictive state $|0\rangle_0 |1\rangle_1 \ldots |1\rangle_N$, associated with detector $D_0$ registering zero photocounts and the other detectors registering one count each, is transformed to the required retrodictive state in input mode 0.

## 2. Predictive state engineering

Before examining the problem of engineering arbitrary retrodictive states of light, it is useful look briefly at a general method for predictive state engineering of travelling optical fields. Any physical state can, to within any required non-zero error, be approximated by truncating it at some finite value $N$ in the photon number state basis:

$$|\psi\rangle = \sum_{n=0}^{N} c_n |n\rangle. \quad (1)$$

A general method for constructing this state is to build it up from the vacuum in a finite number of steps. The simplest way to do this mathematically is by successive application of operators of the type $\hat{E}^\dagger f(\hat{n}) + g(\hat{n})$ to the vacuum where $f(\hat{n})$ and $g(\hat{n})$ are functions of the photon number operator and $\hat{E}^\dagger$ is the bare number raising operator [10]

$$\hat{E}^\dagger = \sum_{n=0}^{\infty} |n+1\rangle\langle n|. \quad (2)$$

It is clear that the presence of the term $g(\hat{n})$ is necessary to ensure that the vacuum state is not removed by the operation. The constructed state is then

$$\left\{ \prod_{n=1}^{N} \left[ \hat{E}^\dagger f_n(\hat{n}) + g_n(\hat{n}) \right] \right\} |0\rangle. \quad (3)$$

Suitable values of $f_n(\hat{n})$ and $g_n(\hat{n})$ can be chosen to ensure that the state (3) is, to within a normalization factor, equal to the desired state $|\psi\rangle$.

A general physical method for realizing (3) for a travelling field has been proposed [7]. In this method the field, which is in a superposition of number states ranging from $|0\rangle$ to $|m\rangle$, is incident on a beamsplitter with a superposition of photon number states $|0\rangle$ and $|1\rangle$ at the other input port. In the output opposite to this latter input is a photodetector. When this detector registers no photons, then the field emerging from the other output is in a superposition of having no photons added and one photon added, that is, it is in a superposition of number states ranging from $|0\rangle$ to $|m+1\rangle$. For this method it is not difficult to show that $f_n(\hat{n}) \propto (\hat{n}+1)^{1/2} t_n^{\hat{n}}$ and $g_n(\hat{n}) \propto t_n^{\hat{n}}$ where $t_n$ is the transmission coefficient of the $n$th beamsplitter. In this way any state that is expressible as a finite superposition of number states can be built up step by step, using a single beamsplitter at each step. In order to obtain the required superposition of photon number states $|0\rangle$ and $|1\rangle$, it was shown how to truncate a coherent state [9], provided a single photon state is available. Thus the building blocks for general quantum state generation by this means are single photon states and coherent states. Dakna et al. [11] have proposed an alternative, but in principle similar, method for constructing a quantum state of a travelling field step by step using single photon states and coherent states.

## 3. Retrodictive state engineering

The particular form of (3) that we shall find useful for this paper is given by $f_n(\hat{n}) = (\hat{n}+1)^{1/2} k^{1/N}$ and $g_n(\hat{n}) = -g_n k^{1/N}$ where $k$ and $g_n$ are complex numbers. By noting that the creation operator $\hat{a}^\dagger$ can be written as $\hat{E}^\dagger (\hat{n}+1)^{1/2}$ we can write expression (3) as

$$k \left[ \prod_{n=1}^{N} (\hat{a}^\dagger - g_n) \right] |0\rangle. \tag{4}$$

In order to express $|\psi\rangle$ given by (1) in the form (4) we write $|n\rangle$ in (1) as $(n!)^{-1/2} (\hat{a}^\dagger)^n |0\rangle$ and equate the resulting expression with (4). Acting on both sides of the resulting equation with $\langle \gamma |$, where $|\gamma\rangle$ is a coherent state, gives [11]

$$k \left[ \prod_{n=1}^{N} (\gamma^* - g_n) \right] = \sum_{n=0}^{N} \frac{c_n}{\sqrt{n!}} (\gamma^*)^n. \tag{5}$$

This expression effectively factorizes the polynomial of degree $N$ on the right-hand side and so the $N$ values of $g_n$ are the $N$ complex roots of the equation in $\gamma^*$ [11]

$$\sum_{n=0}^{N} \frac{c_n}{\sqrt{n!}} (\gamma^*)^n = 0. \tag{6}$$

After substituting the $N$ values of $g_n$ into (4), $k$ can be found as a normalizing constant of the state.

In the multiport device shown in figure 1 there are $N+1$ input ports and $N+1$ output ports with a photodetector in each output port. In $N$ of the input ports there are coherent states $|\beta_m\rangle_m$ with m $= 1, 2 \ldots N$. The detection of any number of photocounts will result in some retrodictive state propagating from the free input port which, without loss of generality, we label as input port 0. For any given photocount pattern, the retrodictive state will be determined by the construction of the device and by the $N$ complex values of $\beta_m$. For versatility, it is useful to have full control over the retrodictive state generated without having to alter the construction of the device. For any state, such as that given by (1), with $N+1$ number-state coefficients there are only $N$ independent coefficients because the other coefficient is determined by normalization. Thus to have the desired control by independent variation of the values of $\beta_m$ we are limited to producing a retrodictive state with $N+1$ number-state coefficients. This limits the total number of photocounts in an acceptable measurement event to $N$. This follows directly from the energy-conserving nature of the device. That is, if a total $N$ photons are detected then the maximum number of photons that can be in input port 0 is $N$. As we are using detectors that only need to distinguish between zero and one photocounts, the detection event we need is that in which all $N$ detectors register one count and the other, which without loss of generality we let be $D_0$, registers no counts.

The multiport device will perform some unitary operation, with an associated unitary operator $\hat{R}$, on the combined input state to produce the combined output state. It is useful to describe these states in terms of creation operators $\hat{a}_n^\dagger$ for each input or output mode $n$ acting on the vacuum. We write the action of the unitary transformation on these operators in the form

$$\hat{R}^\dagger \hat{a}_n^\dagger \hat{R} = \sum_{m=0}^{N} U_{nm}^* \hat{a}_m^\dagger \qquad (7)$$

where $U_{nm}$ are the elements of a unitary matrix.

The measurement event in which each detector registers one count except for $D_0$, which registers zero, corresponds to a probability operator measure (POM) element given by

$$\hat{\Pi}(0) = |0\rangle_{00}\langle 0| \prod_{n=1}^{N} |1\rangle_{nn}\langle 1|. \qquad (8)$$

When this measurement event occurs we assign the retrodictive density operator proportional to $\hat{\Pi}(0)$ [3, 4]. Noting that the trace of $\hat{\Pi}(0)$ is unity, we can write the retrodictive density operator at the time of the measurement as equal to $\hat{\Pi}(0)$, that is as $|\Psi\rangle\langle\Psi|$ where

$$|\Psi\rangle = \left(\prod_{n=1}^{N} \hat{a}_n^\dagger\right)|0\rangle \qquad (9)$$

where $|0\rangle = |0\rangle_0 |0\rangle_1 \ldots |0\rangle_N$ is the combined vacuum state. The backward time evolution of this state through the multiport device yields the retrodictive density operator at the input side of the multiport device as

$$\hat{\rho}^{\text{retr}} = \hat{R}^\dagger |\Psi\rangle\langle\Psi|\hat{R} \qquad (10)$$

where the entangled retrodictive state vector is given by

$$\hat{R}^\dagger |\Psi\rangle = \left(\prod_{n=1}^{N} \hat{R}^\dagger \hat{a}_n^\dagger \hat{R}\right) \hat{R}^\dagger |0\rangle = \left[\prod_{n=1}^{N} \left(\sum_{m=0}^{N} U_{nm}^* \hat{a}_m^\dagger\right)\right] |0\rangle. \qquad (11)$$

Here (7) has been used and also the fact that the vacuum must transform to the vacuum from energy conservation.

We can represent the combined state comprising the predictive density operators for the coherent states $|\beta_m\rangle_m$ in the input modes as a preparation device operator (PDO) [3]

$$\hat{\Lambda}_\beta = \prod_{m=1}^{N} \left(|\beta_m\rangle_{mm}\langle\beta_m|\right). \tag{12}$$

The projection

$$\hat{\rho}_0^{\text{retr}} = \text{Tr}\left(\hat{\rho}^{\text{retr}} \hat{\Lambda}_\beta\right) \tag{13}$$

where the trace is over the $N$ modes $0, 1, \ldots N$ gives the required (unnormalized) retrodictive state. This can be seen by noting that expression (13) is just the time inverse of the projection of a predictive density operator onto an element of a POM representing a measurement of part of the total system. In this latter case the result is an unnormalized predictive state. The PDO given by (12) represents the preparation of part of the system, that is, of the fields in all the input modes except for input mode 0.

Because $\hat{\rho}^{\text{retr}}$ and $\hat{\Lambda}_\beta$ are pure state projectors $\hat{\rho}_0^{\text{retr}}$ will be an unnormalized pure state $|\psi\rangle_{00}\langle\psi|$ with $|\psi\rangle_0$ given by

$$\left(\prod_{m=1}^{N} {}_m\langle\beta_m|\right)\hat{R}^\dagger|\Psi\rangle = \left[\prod_{m=1}^{N}\left({}_m\langle\beta_m|0\rangle_m\right)\right]\left[\prod_{n=1}^{N}\left(U_{n0}^*\hat{a}_0^\dagger + \sum_{m=1}^{N} U_{nm}^*\beta_m^*\right)\right]|0\rangle_0. \tag{14}$$

This is of the form (4) where $g_n$ for $n = 1, 2, \ldots, N$ is given by

$$g_n = -\frac{1}{U_{n0}^*}\sum_{m=1}^{N} U_{nm}^*\beta_m^* \tag{15}$$

and $k$ is replaced by

$$\bar{k} = \exp\left(-\frac{1}{2}\sum_{m=1}^{N}|\beta_m|^2\right)\prod_{n=1}^{N} U_{n0}^*. \tag{16}$$

We note that here $\bar{k}$, as opposed to $k$, is not a normalizing constant.

In order to engineer the retrodictive state whose state is given by (4) we need to determine the appropriate phase and amplitude of each of the $N$ coherent states, that is, we need to determine the complex numbers $\beta_m$ in terms of $g_n$ for $n = 1, 2, \ldots, N$. To do this we first note that (15) does not represent a unitary transformation, because it does not involve all the elements of $U_{nm}$. To overcome this, we define additional quantities $\beta_0 = 0$ and

$$g_0 \equiv -\frac{1}{U_{00}^*}\sum_{m=0}^{N} U_{0m}^*\beta_m^*. \tag{17}$$

We can then combine (15) and (17) to give

$$-U_{n0}^* g_n = \sum_{m=0}^{N} U_{nm}^*\beta_m^* \quad \text{for } n = 0, 1, \ldots, N, \tag{18}$$

Retrodictive quantum optical state engineering    1619

which is a unitary transformation and can be easily inverted to yield

$$\beta_m^* = -\sum_{n=0}^{N} U_{nm} U_{n0}^* g_n \quad \text{for } m = 0, 1, \ldots, N. \qquad (19)$$

By solving (19) with $\beta_0 = 0$ we find that the additional variable $g_0$ can be written as

$$g_0 = -\frac{1}{|U_{00}|^2} \sum_{n=1}^{N} |U_{n0}|^2 g_n. \qquad (20)$$

Thus given a lossless multiport device with $N+1$ inputs and the same number of outputs with coherent states in $N$ of the input ports and detectors in the all the output ports capable of distinguishing one from zero photocounts, *any* retrodictive state that can be approximated by an expansion in the first $N+1$ photon number states can in general be generated. The coherent states needed to generate a particular state can be determined by the following two steps. The first step is to express the state to be generated, which is in the number state superposition (1), in the form (4). To do this, the $N$th degree equation (6) is solved to give the $N$ values of $g_n$ for $n=1, 2, \ldots, N$ and $g_0$ is obtained from (20). The second step is simply to find the required values of $\beta_m$ for $m=1, 2, \ldots, N$ from (19).

## 4.   Optimization

In the above method for retrodictive state engineering we assumed that the multiport device, and thus the unitary transformation, was fixed and the state generated was controlled by the coherent state inputs. It is also possible, however, to control the generated state by altering the multiport device. Reck *et al.* [12] have shown how it is possible to construct a multiport device from beamsplitters, mirrors and phase shifters to realize *any* unitary transformation. The construction is not unique and many unitary transformations can be realized by a variety of configurations. In most practical situations, however, it would probably be easier to adjust the coherent state inputs than the multiport device, once the latter has been set up. In general different unitary transformations will lead to different probabilities of success in engineering a particular retrodictive state. In this section we shall consider a single unitary transformation that should have a good applicability to a reasonably wide range of states. We shall also consider how to optimize the transformation for a particular state and investigate replacing the $N$ coherent state inputs by $N-1$ vacuum inputs plus a single non-vacuum coherent state input.

### 4.1.   *Versatile multiport configuration*

In finding a versatile general unitary transformation that can be used to generate a large number of retrodictive states with a reasonable success probability, it is useful first to consider unitary transformations that are not very useful. The simplest of these is the unit operator. This links input 0 only with output 0, input 1 only with output 1 and so on. Clearly with output detectors that can distinguish one from zero photons the only retrodictive states that can be generated at input 0 with such a transformation are the vacuum and one-photon states. To generate a superposition of a larger number of photon number states, we need more extensive linkages. When some of the linkages vanish, the generation of some particular retrodictive states will be impossible. The most extensive linkage system is where a photon incident in any input has a non-zero probability of being detected at any output and a photon detected at any output has a non-zero probability of having entered at any input. When some of these linkages are non-zero but very weak, we might expect that the probability of generating some particular retrodictive states is very small. The most versatile multiport device would therefore seem to be one in which an input photon has the same probability of being detected at any output and a detected photon has the same probability of having entered any input. Such devices, which can be considered as generalizations of the 50:50 symmetric beamsplitter, have been studied in [13]. A particularly useful example of such a device is one that performs a discrete Fourier transform [14]. The elements of the unitary transformation matrix for this device can be written simply as

$$U_{nm} = \frac{\omega^{nm}}{\sqrt{N+1}} \qquad (21)$$

where $\omega = \exp[i2\pi/(N+1)]$ is a $(N+1)$th root of unity. The set of transformed operators form a discrete Fourier transform pair. An example with four input and four output ports is the eight-port interferometer [15] as used in [16]. A different example of a Fourier transform multiport with four input and four output ports is illustrated in [17], which shows that the construction of such a device is not unique.

### 4.2. *Optimization for particular state*

While the above multiport device will have application to a wide range of states, in some cases a particular state, or states obtainable from it by, for example, a simple phase shift, is all that is required. In this case it is worth optimizing the transformation for the particular state needed. The quantity we wish to maximize is the probability that the retrodictive state will be generated, that is, we wish to maximise the probability of the detection event in which detector $D_0$ detects zero photons while the other detectors detect one photon each. To do this we note that just as the suitably normalized POM element for this event can be used to assign the retrodictive state at the time of detection, so too can a retrodictive density operator be used as a measurement device operator (MDO) [3]. As $\hat{\rho}_0^{\text{retr}}$ is the retrodictive state in mode 0 at the input port 0, this will be a MDO for the measurement device comprising all of figure 1 except for input port 0. That is, the measurement device that includes the detectors, the multiport device and all the fields in the $N$ input ports other than input port 0. If we label this MDO as $\hat{\Gamma}_0$, then the probability of the required event for any predictive field in input port 0 is proportional to the trace of the product of the density operator for this field and $\hat{\Gamma}_0$. This in turn is proportional to $|\bar{k}|^2$ where $\bar{k}$ is given by (16). Thus it is this quantity that should be maximised.

In determining the value of $|\bar{k}|^2$ it is necessary to evaluate the sum of $|\beta_m|^2$. From (19) we find, making use of the unitary nature of the transformation matrix, that

$$\sum_{m=0}^{N} |\beta_m|^2 = \sum_{m=0}^{N} |U_{m0}|^2 |g_m|^2 \qquad (22)$$

which gives

$$|\bar{k}|^2 = \exp\left(-\sum_{m=0}^{N}|U_{m0}|^2|g_m|^2\right)\prod_{n=1}^{N}|U_{n0}|^2. \qquad (23)$$

The values of $|g_m|^2$ for $m \neq 0$ are determined by the retrodictive state that we wish to generate and $|g_0|^2$ is a function of these and $U_{n0}$, so to maximize $|\bar{k}|^2$ there is no need to vary the coherent state amplitudes independently. Given the state we wish to generate, we can optimize its generation probability by varying the moduli of the $N+1$ transformation matrix elements $U_{n0}$ subject to the constraint

$$\sum_{n=0}^{N}|U_{n0}|^2 = 1. \qquad (24)$$

Also from (15) $U_{n0}$ must not be zero. In general this maximization can be achieved using Lagrange's undetermined multipliers.

### 4.3. *Single coherent state input*

It is interesting to note that, for a given state to be generated, $|\bar{k}|^2$ in (23) depends only on $|U_{n0}|$. Thus maximizing $|\bar{k}|^2$ for a particular retrodictive state affects only the first column of the transformation matrix. From (22) this then determines the sum of $|\beta_m|^2$. There is scope, therefore, to choose the arguments of $U_{n0}$, for example on the basis of simplicity of physical construction of the multiport, and then to adjust the remaining elements of the matrix to allow convenient individual values of $|\beta_m|$. A useful configuration would be one in which we need only a *single* input coherent state of non-zero amplitude with vacuum states in the other inputs. To be definite, we let $\beta_m = 0$ except for $\beta_1$ and assume we have chosen the first column elements $U_{n0}$. Then from (18) the elements of the second column are given by

$$U_{n1} = -\frac{g_n^* U_{n0}}{\beta_1} \qquad \text{for } n = 0, 1, \ldots, N. \qquad (25)$$

We can find $|\beta_1|$ in terms of the elements of the first column of the transformation matrix from (22) and (20). From (25) this then determines the moduli of $U_{n1}$ that are needed to generate the required retrodictive state with a single coherent state input. The arguments of these elements depend, through the relation (25), on the phase of the coherent state input. This phase can be chosen arbitrarily and the matrix elements matched to it. It is not difficult to show that the conditions on $U_{n1}$ necessary to maintain the unitarity of the transformation, that is

$$\sum_{n=0}^{N}|U_{n1}|^2 = 1 \qquad (26)$$

$$\sum_{n=0}^{N}U_{n0}^* U_{n1} = 0 \qquad (27)$$

are then satisfied.

The above shows that it is always possible to reduce the number of non-vacuum coherent state inputs to just one. Further, a multiport device with a single non-vacuum coherent state input will give the same optimum performance, that is maximum $|\bar{k}|^2$, as a similar device with multiple coherent state inputs. The single coherent state input case can be considered as the natural method of altering the original projection synthesis device of reference [6] to a function with an easily prepared coherent state input in place of the exotic reciprocal binomial state. This is achieved by replacing the original beamsplitter by a multiport device. If vacuum states are in $N-1$ input ports of the multiport, only two input ports need to be accessible. This allows the possible use of two modified plate beamsplitters for the construction of the multiport as illustrated in reference [18].

## 5. Retrodictive phase states

By way of an example of our general approach, we consider the generation of some retrodictive truncated phase states. These will be useful, for example, for measuring the phase distribution of weak states of light by projection synthesis. For very weak fields, which must have broad phase distributions, homodyne methods do not yield the canonical phase distribution [19]. For such weak fields a retrodictive phase state truncated at a small photon number is sufficient.

The simplest non-trivial retrodictive phase state is proportional to $|0\rangle + \exp(i\theta)|1\rangle$. To generate this it is sufficient to generate the state proportional to $|0\rangle + |1\rangle$ and allow it to evolve backwards in time through an appropriate phase shifter. The first step is to convert $|0\rangle + |1\rangle$ to the form (4). As $N=1$ there is just one term involved, giving $g_1 = -1$. Then from (20)

$$g_0 = \frac{|U_{10}|^2}{|U_{00}|^2}. \tag{28}$$

From (19) we then find

$$\beta_1^* = U_{11} U_{10}^* - U_{01} U_{00}^* g_0. \tag{29}$$

We shall consider first generating this retrodictive state with a discrete Fourier transform device with two inputs and two outputs. This can be constructed from a beamsplitter with a $\pi/2$ phase shifter in the input and output ports of mode 1 or from a Mach-Zehnder interferometer. Here $\omega = -1$ so from (21)

$$U_{nm} = \frac{(-1)^{nm}}{\sqrt{2}} \tag{30}$$

and we find easily that the coherent state input $|\beta_1\rangle_1$ must be such that $\beta_1 = -1$. From (16) we find that $|\bar{k}|^2 = 0.184$. This compares with the normalizing constant in (4) for which in this case $k^2 = 0.5$. Thus when the generated retrodictive state is used as a POM element for a phase measurement, the probability of a successful detection event is 37% of that for the case where the POM element is the normalized truncated phase state projector.

To maximize $|\bar{k}|^2$ by adjusting the transformation matrix we find from (23) that

$$|\bar{k}|^2 = |U_{10}|^2 \exp\left(-|U_{10}|^2 - |U_{10}|^4 |U_{00}|^{-2}\right). \tag{31}$$

Substituting $|U_{10}|^2 = 1 - |U_{00}|^2$ from (24) gives a single-variable expression which is straightforward to maximize. We find the maximum value of $|\bar{k}|^2$ is 0.206 which occurs when $|U_{00}|^2 = 0.618$. For a beamsplitter this corresponds to a ratio $|t|^2 : |r|^2$ of 62:38 where $t$ and $r$ are the transmission and reflection coefficients. This optimization increases the 37% probability ratio obtainable with a Fourier transform multiport to 41%. In practice such a small increase may not warrant the alteration of the multiport from the Fourier transform configuration. The coherent state input $|\beta_1\rangle_1$ must now be such that $|\beta_1| = 0.786$ with a phase depending on the phases chosen for the matrix elements. These results are in accord with those of Clausen *et al.* [20] who considered the measurement of the overlap of a quantum state with a finite superposition of Fock states by means of a chain of beamsplitters with the same transmission and reflection coefficients, with the single beamsplitter being treated as a simple example of this chain. Such a chain is, of course, a particular case of the general lossless multiport device considered here.

The second case we examine is the generation of the retrodictive state $|0\rangle + |1\rangle + |2\rangle$. As $N=2$, equation (6) is quadratic, with solutions

$$g_1 = -\frac{1}{\sqrt{2}} - i\sqrt{\sqrt{2} - 1/2} \tag{32}$$

$$g_2 = g_1^*. \tag{33}$$

The Fourier transform multiport for this case has matrix elements $U_{nm} = 3^{-1/2} \omega^{nm}$ with $\omega = \exp(i2\pi/3)$ so, from (19), we obtain

$$\beta_1^* = -(g_0 + \omega g_1 + \omega^2 g_2)/3 \tag{34}$$

$$\beta_2^* = -(g_0 + \omega^2 g_1 + \omega g_2)/3 \tag{35}$$

which from (20) yields $\beta_1 = -1.259$ and $\beta_2 = -0.155$ for the required input coherent states. This gives $|\bar{k}|^2 = 0.022$ compared with $k^2 = 1/6$, so the probability ratio is 13.3%.

The maximization of $|\bar{k}|^2$ involves finding the maximum of a function of two variables. We outline a procedure for this in the Appendix based on Lagrange's undetermined multipliers which can be extended to higher values of $N$. We find $|U_{00}|^2 = 0.436$ and $|U_{10}|^2 = |U_{20}|^2 = 0.282$. The maximum value of $|\bar{k}|^2$ is 0.0248, which increases the probability ratio to 14.9%. Again we see that the optimization gives only a small improvement over the Fourier transform configuration. For this maximum value, $|\beta_1|^2 + |\beta_2|^2 = 1.162$. By selecting second column elements of the transformation matrix such that $|U_{01}|^2 = 0.314$ and $|U_{11}|^2 = |U_{21}|^2 = 0.343$, this can be achieved with a single coherent state input with $|\beta_1| = 1.078$.

## 6. Conclusion

In this paper we have examined the engineering of retrodictive states of light by means of a general lossless multiport device. We have restricted the input predictive states to be coherent states because of their relative ease of preparation. We have also restricted the photodetectors in the outputs to those that can distinguish between zero and one photons. With such detectors the widest range of retrodictive states are produced when all the detectors register one photocount, except one which registers zero counts. This is the case we study here. We should point out, however, while this photocount pattern gives the broadest range of states, useful particular retrodictive states can also be produced when other photocount patterns are registered as, for example, in reference [8]. The beamsplitter chain suggested in reference [20], if examined from the perspective of retrodictive state engineering, can be seen to be a particular example of the general multiport device considered in this paper.

We have not considered the effects of detector inefficiency, which in general will result in the generated retrodictive state being mixed rather than pure. The final effect of the inefficiency will depend on the use to which the retrodictive state is put. For example, if it is for the generation of a predictive state, as in the quantum scissors device [9], this predictive state will also be mixed. If the retrodictive state is used to measure the probability distribution of an optical observable, such as phase, then the measured probabilities that give the distribution can be corrected for detector imperfections [8, 21].

The input coherent states give control of the retrodictive state produced but whether or not the state is produced successfully depends on the probability of obtaining the appropriate photocount pattern. We have shown how the multiport configuration can be optimized to produce the greatest probability of a successful experiment and how the number of non-vacuum coherent state inputs can be reduced to just one without affecting this optimum probability. Changing the multiport will in general not be as simple as changing the coherent state inputs, so a versatile multiport arrangement capable of producing a range of retrodictive states with reasonable probability is useful. In this regard we have examined in particular a multiport configuration whose transformation corresponds to a discrete Fourier transform. This may be considered as the natural generalization of the 50:50 beamsplitter. We have shown that to produce some retrodictive phase states there is not a huge difference in success probability between the Fourier transform and the optimal configurations.

An interesting question arises as to whether or not the general multiport device discussed here for retrodictive state engineering can also be used for predictive state engineering simply by reversing the roles of preparation and measurement. The inverse situation to that described in this paper is when there are $N$ single photon states plus a vacuum at the $N+1$ input ports and coherent states are detected at $N$ of the outputs, with the required predictive state being generated at the free output. While the intrinsic symmetry of quantum mechanics allows this, it is technically far more difficult than the retrodictive case. Producing single photon states on demand and constructing coherent state POM elements for the measurement is far more difficult than producing coherent states on demand and detecting single photons. Thus we have the perhaps surprising result that retrodictive state engineering, at least by means of the general technique described here, is much easier than predictive state engineering.

**Acknowledgement**

DTP thanks the Australian research council for funding.

**Appendix.  Optimization for $N=2$**

Here we outline a procedure based on Lagrange's undetermined multipliers for determining the optimal multiport configuration for the generation of the retrodictive state $|0\rangle + |1\rangle + |2\rangle$.

Writing $x_n = |U_{n0}|^2$ for $n = 1, 2, 3$, we find from (23) that the expression we have to maximise is

$$|\bar{k}|^2 = x_1 x_2 \exp\left[ -\frac{\sqrt{2}(x_1^2 + x_2^2) + 2x_1 x_2(1 - \sqrt{2})}{x_0} - \sqrt{2}(x_1 + x_2) \right] \qquad (A1)$$

subject to the constraint

$$x_0 + x_1 + x_2 = 1 \tag{A2}$$

The condition for a maximum, $d|\bar{k}|^2 = 0$, can be combined with the constancy of (A2) to give

$$d|\bar{k}|^2 + \lambda d(x_0 + x_1 + x_2) = 0 \tag{A3}$$

where $\lambda$ is an undetermined multiplier. From this we may write

$$\frac{\partial |\bar{k}|^2}{\partial x_0} + \lambda = 0 \tag{A4}$$

$$\frac{\partial |\bar{k}|^2}{\partial x_1} + \lambda = 0 \tag{A5}$$

$$\frac{\partial |\bar{k}|^2}{\partial x_2} + \lambda = 0 \tag{A6}$$

Finding the partial derivatives and subtracting (A6) from (A5) yields $x_1 = x_2$ and thus $x_0 = 1 - 2x_1$. Subtracting (A5) from (A4) and then substituting for $x_0$ gives eventually a cubic equation in $x_1$ of which the only real solution is $x_1 = 0.282$ and thus $x_0 = 0.436$. These values, with $x_1 = x_2$, give the maximum value of $|\bar{k}|^2$ as 0.0248.

## References
[1] WATANABE, S., 1955, *Rev. Mod. Phys.*, **27,** 179.
[2] AHARONOV, Y., BERGMAN, P. G., and LEBOWITZ, J. L., 1964, *Phys. Rev.*, **134,** B1410; PENFIELD, R. H., 1966, *Am. J. Phys.,* **34,** 422; AHARONOV, Y., and ALBERT, D. Z., 1984, *Phys. Rev.* D, **29,** 223; AHARONOV, Y., and ALBERT, D. Z., 1984, *Phys. Rev.* D, **29,** 228; AHARONOV, Y., and VAIDMAN, L., 1991, *J. Phys. A: Math. Gen.,* **24,** 2315.
[3] BARNETT, S. M., PEGG, D. T., and JEFFERS, J., 2000, *J. Mod. Opt.*, **47,** 1779; PEGG, D. T., BARNETT, S. M., and JEFFERS, J., 2002, *J. Mod. Optics*, **49,** 913; PEGG, D. T., BARNETT, S. M., and JEFFERS, J., 2002, *Phys. Rev.* A, **66,** 022106.
[4] PEGG, D. T., and BARNETT, S. M., 1999, *J. Opt. B: Quantum Semiclass. Opt.*, **1,** 442; BARNETT, S. M., PEGG, D. T., JEFFERS, J., and JEDRKIEWICZ, O., 2001, *Phys. Rev. Lett.,* **86,** 2455; JEFFERS, J., BARNETT, S. M., and PEGG, D. T., 2002, *J. Mod. Opt.*, **49,** 925; JEFFERS, J., BARNETT, S. M., and PEGG, D. T., 2002, *J. Mod. Opt.*, **49,** 1175; CHEFLES, A., and SASAKI, M., 2003, *Phys. Rev.* A, **67,** 032112.
[5] BARNETT, S. M., PEGG, D. T., JEFFERS, J., JEDRKIEWICZ, O., and LOUDON, R., 2000, *Phys. Rev.* A, **62,** 022313.
[6] BARNETT, S. M., and PEGG, D. T., 1996, *Phys. Rev. Lett.*, **76,** 4148; PHILLIPS, L. S., BARNETT, S. M., and PEGG, D. T., 1998, *Phys. Rev.* A, **58,** 3259.
[7] PEGG, D. T., BARNETT, S. M., and PHILLIPS, L. S., 1997, *J. Mod. Opt.*, **44,** 2135.
[8] PREGNELL, K. L., and PEGG, D. T., 2001, *J. Mod. Opt.*, **48,** 1293; PREGNELL, K. L., and PEGG, D. T., 2002, *J. Mod. Opt.*, **49,** 1135; PREGNELL, K. L. and PEGG, D. T., 2002, *Phys. Rev.* A, **66,** 013810.
[9] PEGG, D. T., PHILLIPS, L. S., and BARNETT, S. M., 1998, *Phys. Rev. Lett.,* **81,** 1604; BARNETT, S. M., and PEGG, D. T., 1999, *Phys. Rev.* A, **60,** 4965.
[10] SUSSKIND, L., and GLOGOWER, J., 1964, *Physics*, **1,** 49.
[11] DAKNA, M., CLAUSEN, J., KNÖLL, L., and WELSCH, D.-G., 1999, *Phys. Rev.* A, **59,** 1658.

[12] RECK, M., ZEILINGER, A., BERNSTEIN, H. J., and BERTANI, P., 1994, *Phys. Rev. Lett.*, **73,** 58.
[13] MATTLE, K., MICHLER, M., WEINFURTER, H., ZEILINGER, A., and ZUKOWSKI, M., 1995, *Appl. Phys.* B, **60,** S111.
[14] TÖRMÄ, P., STENHOLM, S., and JEX, I., 1995, *Phys. Rev.* A, **52,** 4853.
[15] WALKER, N. G., and CARROLL, J. E., *J. Mod. Opt.*, **34,** 15; NOH, J.W., FOUGÈRES, A., and MANDEL, L., 1991, *Phys. Rev. Lett.*, **67,** 1426.
[16] PREGNELL, K. L., and PEGG, D. T., 2003, *Phys. Rev.* A, **67,** 063814.
[17] PREGNELL, K. L., and PEGG, D. T., 2002, *Phys. Rev. Lett.*, **89,** 173601.
[18] TÖRMÄ, P., and JEX, I., 1996, *J. Mod. Opt.*, **43,** 2403.
[19] VACCARO, J. A., and PEGG, D. T., 1994, *Opt. Commun.*, **105,** 335.
[20] CLAUSEN, J., DAKNA, M., KNÖLL, L., and WELSCH, D.-G., 2000, *Opt. Commun.,* **179,** 189.
[21] LEE, C. T., 1993, *Phys. Rev.* A, **48,** 2285.